\documentclass[%
aps,
pre,
preprint,
groupedaddress,
amsmath,amssymb,
]{revtex4-1}

\usepackage{graphicx}
\usepackage[squaren,Gray]{SIunits}
\usepackage{chemist}

\usepackage{xcolor, soul}
\sethlcolor{cyan}

\begin{document}

\title{Fluctuations of a membrane nanotube covered with an actin sleeve}

\author{A. Allard}
\email{antoine.allard@curie.fr}
\altaffiliation{Also at LAMBE}
\author{F. Valentino}
\email{valentino@crans.org}
\author{C. Sykes}
\email{cecile.sykes@curie.fr}
\affiliation{Laboratoire Physico Chimie Curie, Institut Curie, PSL Research University, CNRS UMR168, Paris, France}
\affiliation{Sorbonne Universités, UPMC Univ. Paris 06, Paris, France}
\author{T. Betz}
\email{timo.betz@uni-muenster.de}
\affiliation{Institute of Cell Biology, Cells in Motion Interfaculty Center, Centre for Molecular Biology of Inflammation, Münster, Germany}
\author{C. Campillo}
\email{clement.campillo@univ-evry.fr}
\affiliation{LAMBE, Université d'Evry, CNRS, CEA, Université Paris-Saclay, 91025, Evry-Courcouronnes, France}

\date{\today}

\begin{abstract}
Many biological functions rely on the reshaping of cell membranes, in particular into nanotubes, which are covered \textit{in vivo} by dynamic actin networks. Nanotubes are subject to thermal fluctuations, but the effect of these on cell functions is unknown. Here, we form nanotubes from liposomes using an optically trapped bead adhering to the liposome membrane. From the power spectral density of this bead, we study the nanotube fluctuations in the range of membrane tensions measured \textit{in vivo}. We show that an actin sleeve covering the nanotube damps its high frequency fluctuations because of the network viscoelasticity. Our work paves the way for further studies on the effect of nanotube fluctuations in cellular functions.
\end{abstract}


\maketitle

Living organisms are dynamic systems which constantly adapt their morphology. Their shape changes rely on the remodeling of the lipid membranes that delineate cell boundaries as well as intracellular compartments. Inside the cell, membranes are often found in narrow tubules which are cylinders made of a single lipid bilayer, here referred as nanotubes \cite{renard2015}. For example, some tubules are transient, like the ones extruded from the plasma membrane or from the Golgi apparatus \cite{Miserey-Lenkei2010}, while some other tubular structures have a permanent cylindrical shape, such as the tubular network of the endoplasmic reticulum (ER), a complex organelle extended all over the cell from the vicinity of the nucleus towards the cell membrane \cite{park2010}. The ER is thus made of interconnected nanotubes fluctuating at 0.1 to 1 second time scale, thus making the whole organelle highly dynamic \cite{nixon2016,Georgiades2017}. Despite the high dynamics, the effect of these fluctuations on nanotube fates is unknown. Moreover, actin networks directly interact with nanotubes in the cell \cite{prinz2000,kaksonen2006,romer2010,Miserey-Lenkei2010,echarri2012,dai2019}, but the mechanical effect of this interaction also remains unclear. In this article, we assess nanotube fluctuations at membrane tension (\unit{0.2-50\times10^{-6}}{\newton\per\meter}) similar to \textit{in vivo} situations and in the presence of an actin network. This approach is inspired by experimental and theoretical work on membrane fluctuations \cite{evans1986,Fournier2007,betz2012,Valentino2016,turlier2016,barooji2016,Mirzaeifard2016}.

Nanotubes spontaneously extrude from a plane membrane upon application of a well characterized pulling force \cite{Waugh1987,derenyi2002}. This force depends on the membrane tension $\sigma$ which ranges \textit{in vivo} from \unit{5\times10^{-6}}{\newton\per\meter} for the Golgi membrane to \unit{13\times10^{-6}}{\newton\per\meter} for ER membrane \cite{Upadhyaya2004}. Here, we extrude nanotubes from settled and slightly adherent liposomes using a bead held in an optical trap \cite{Allard}. We access the nanotube fluctuations through the power spectral density (PSD) of the trapped bead connected to the nanotube. Indeed, our setup allows us accessing bead position at a high spatial ($\unit{1}{\nano\meter}$) and temporal ($\unit{4}{\micro\second}$) resolution (Fig. \ref{fgr:setup}(a)) \cite{gittes1997,tolic2006,vermeulen2006,Valentino2016}. Our \textit{in vitro} assay allows us to control the properties of both, membrane and actin network, thus avoiding the complexity of the cell interior.

We show that the presence of a membrane nanotube at low tension increases the PSD of the bead (in the absence of the nanotube) in the frequency regime \unit{[1-100]}{\hertz}. We explain this increase using our previous model that predicts a shift of the frequency regime where peristaltic undulations of the nanotube dominate bead fluctuations \cite{Valentino2016}.

Then, we compare the bead PSD before and after actin polymerization. For frequencies between 0.5 and \unit{5}{\kilo\hertz}, the PSD is described by a power law whose exponent increases in the presence of the actin network whereas the amplitude of the corresponding fluctuations decreases. Those observations stem from the viscoelasticity of the actin architecture that we include in our theoretical framework. Indeed, the grown actin network behaves as a viscoelastic material \cite{tseng2001,juelicher2007,Gardel2008}. Therefore, we demonstrate that actin modulates the local undulation of membrane nanotubes. This could play a role \textit{in vivo} on the stability of membrane tubules and their interactions with membrane remodeling proteins.

\section*{Results}

\subsection*{Experimental assay}

    Membrane nanotubes are obtained by first trapping a polystyrene bead that specifically binds to biotinylated lipids (Materials and methods in Supplementary materials). To extrude a nanotube at low membrane tension, we use liposomes slightly adhering on a substrate and then move the stage away. Measuring the nanotube force and knowing the membrane bending energy $\kappa$, we infer the tension $\sigma = F^2/8\pi^2\kappa$ \cite{Waugh1987,derenyi2002}. In our conditions, tension ranges between 0.2 and \unit{50\times10^{-6}}{\newton\per\meter}, while aspirating liposomes in a micropipette gives \unit{10-200\times10^{-6}}{\newton\per\meter} \cite{Valentino2016,Allard}. The detailed effect of membrane tension is assessed in section \textit{Temporal nanotube fluctuations at low tension}.
    
\begin{figure}[b]
    \centering
        \includegraphics{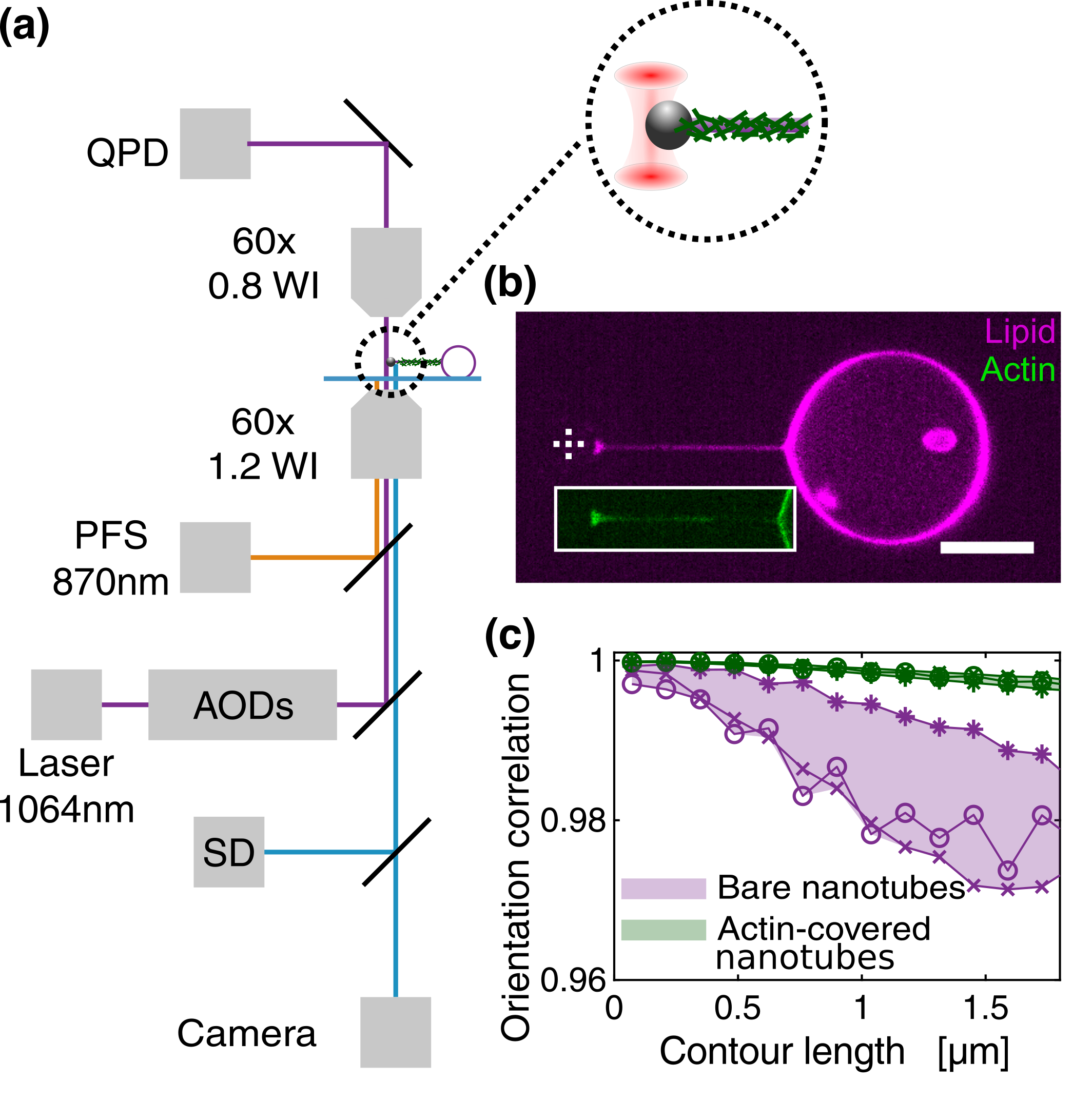}
        \caption{\textbf{Experimental setup} (a) Liposomes are settled down on a glass surface and imaged by a spinning disk confocal microscope (SD) and a camera. Beads are optically trapped using an infrared \unit{1064}{\nano\meter} laser coupled with an AOD pair and injected into the optical path. The laser signal is collected in transmission on a quadrant-photodiode (QPD). A perfect focus system (PFS) maintain the focus on live. For clarity, lenses and micropipettes are not represented. Insert on the right, magnified scheme of a grown actin network (green) at the surface of a membrane nanotube (magenta). (b) Confocal image of a membrane nanotube covered by an actin network. Lipid in magenta and actin in green. Dashed cross indicates bead center. Scale bar \unit{10}{\micro\meter}. (c) Orientation correlation function as a function of the contour length along the nanotube, calculated using the Easyworm software \cite{Lamour2014}. Three bare nanotubes (star, circle and cross symbols) display an orientation correlation function under one (magenta). Once covered with actin, the correlation goes to one (green).}
    \label{fgr:setup}
\end{figure}

    To decorate the membrane with actin, we polymerize a branched actin network at the surface of the membrane nanotube in a two-step procedure \cite{Allard}. First, we specifically bind pVCA to the nanotube which further activates actin polymerization (Materials and methods in Supplementary material). In a second step we supply actin monomers to the nanotube and thus an actin sleeve forms at the nanotube surface (Fig. \ref{fgr:setup}(a and b)).
    
    To first assess whether the presence of an actin sheath on nanotubes could affect their fluctuations, nanotubes are imaged at a rate of 1 frame per second with a spinning disc confocal microscope before and after actin polymerization. The shape of these nanotubes is extracted over time from their lipid signals, and their local orientation is measured using an open-source Matlab code \cite{Lamour2014}. The orientation correlation function is determined along the contour length of the nanotube in the presence and in the absence of an actin sleeve. The orientation correlation function, is given by: $C(\ell)=<\cos(\theta(s;s+\ell))>_{s,t}$, where $\theta(s;s+\ell)$ is the angle between the tangents at the curvilinear abscissae $s$ and $s+\ell$ at a given time. The cosine is averaged over all curvilinear abscissae $s$ and over time $t$. $C(\ell)$ quantifies whether the shape of the nanotube is linear or not: theoretically $C(\ell)=1$ corresponds to perfectly straight nanotubes, whereas $C(\ell)<1$ is associated with curved nanotubes.
    
    We have studied 20 independent nanotubes. In the absence of actin, most of them (N = 17/20) appear to be straight, thus $C(\ell<\unit{2}{\micro\meter})>0.99$. We have chosen the three nanotubes exemplified in Fig. \ref{fgr:setup}(c) that have a value of $C$ lower than 0.99 for $\ell=\unit{2}{\micro\meter}$, therefore showing measurable fluctuations (magenta stars, circles and crosses in Fig. \ref{fgr:setup}(c) and Supplementary Movie). Then, we observed that these fluctuations disappear in the presence of an actin sleeve ($C(\ell<\unit{2}{\micro\meter})>0.99$, green stars, circles and crosses in Fig. \ref{fgr:setup}(c)). Note that for the 17 remaining nanotubes, no significant differences are observed, compared to the initial situation (bare nanotubes). These indicate that the presence of an actin sleeve reduces membrane nanotube spatial undulations observed at a rate of $\unit{1}{\hertz}$ and motivates a closer look in a larger range of frequencies ($\unit{1}{\hertz}$ - $\unit{25}{\kilo\hertz}$). To do so, we record the fluctuations of the bead connected to the nanotube to explore its fluctuations amplitude as a function of the frequency.
    
\begin{figure*}[b]
  \centering
    \includegraphics[width=\textwidth]{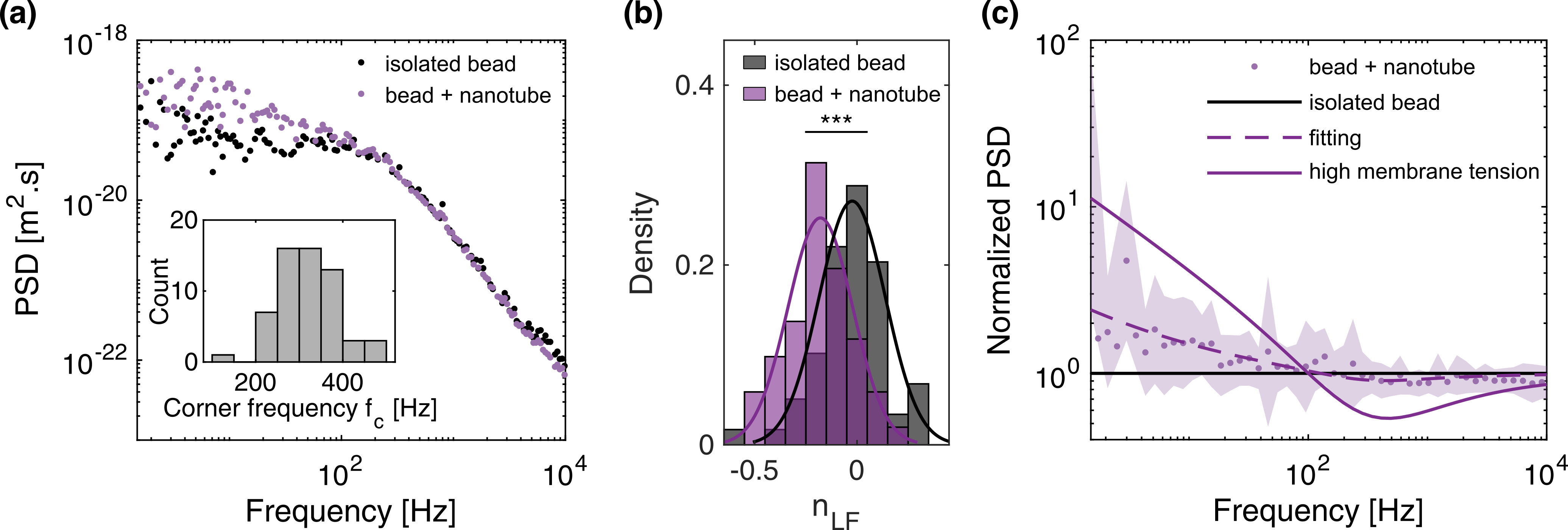}
    \caption{(a) Power Spectral Densities (PSDs) as a function of the frequency $f$ for an isolated bead (black) and the same bead used to pull a bare membrane nanotube (magenta). Inset: Distribution of the corner frequency $f_{\rm{c}}$ for N = 59 beads (Materials and methods in Supplementary material). (b) PSD exponent within a low frequency range (\unit{[10-100]}{\hertz}) for isolated beads (black, N = 59) and beads connected to a bare membrane nanotube (magenta, N = 51). p-values calculated using Student t-test. *** p $<$ 0.001. (c) Ratio between the ${\rm PSD_t}$ of a bead connected to a membrane nanotube (equation $\eqref{eq:PSDV}$) and the ${\rm PSD_b}$ of the same isolated bead (equation $\eqref{PSDfree}$). Magenta dots represent the statistic average ratio overall nanotube (N = 51) and the magenta region is its standard deviation. Magenta dash line fits the average data against $f$, with $f_{\rm c} = \unit{320}{\hertz}$ and $f_{\rm t}$ as a free parameter. This gives $f_{\rm{t}} = \unit{3.1\pm3.5}{\hertz}$. Dark line indicates the predicted curve for an isolated bead ($f_{\rm{t}} = \unit{0}{\hertz}$). Magenta line indicates the prediction for nanotube pulled from liposomes under pressure as in \cite{Valentino2016} ($f_{\rm{t}} = \unit{200}{\hertz}$).}
    \label{fgr:PSDTube}
\end{figure*}

\subsection*{Temporal nanotube fluctuations at low tension}
    The fluctuations of a micrometric bead are captured by the measure of its power spectral density (PSD, equation $\eqref{PSDdef}$). In the case of an ``isolated bead'', the bead is held by the optical trap and fluctuates because of the thermal agitation in the surrounding viscous fluid. The bead undergoes an elastic force, a Brownian force and Stokes force. The Fourier transform of the Langevin equation, reflecting the bead dynamics gives the theoretical PSD of the fluctuating trapped bead \cite{gittes1997}:
    
    \begin{equation}
        {\rm PSD}_{\rm{b}}(f) = \frac{k_{\rm{B}}T}{12\pi^3\eta r_{\rm{bead}}}\frac{1}{f_{\rm{c}}^2+f^2}
    \label{PSDfree}
    \end{equation}
    where $k_{\rm{B}}$ is the Boltzmann constant, $T$ the bath temperature, $\eta$ the viscosity of the surrounding fluid, $r_{\rm{bead}}$ the bead radius and $f_{\rm{c}}$ the corner frequency that reflects the optical trap stiffness and is given by:
    
    \begin{equation}
        f_{\rm{c}} = \frac{k_{\rm{trap}}}{12\pi^2\eta r_{\rm{bead}}}
        \label{eq:fc}
    \end{equation}
    with $k_{\rm{trap}}$ the trap stiffness. Fig. \ref{fgr:PSDTube}(a) shows that equation $\eqref{PSDfree}$ accurately describes the experimental PSD of the bead in the optical trap, and we measure $f_{\rm{c}} = \unit{320 \pm 70}{\hertz}$ for different beads (mean $\pm$ st.d., N = 59, inset Fig. \ref{fgr:PSDTube}(a)).
    
    The PSD exhibits two distinct regimes with $f_{\rm c}$ as a corner frequency: for $f\ll f_{\rm c}$, equation $\eqref{PSDfree}$ states that ${\rm PSD}_{\rm{b}}\sim f^0$ is frequency-independent whereas for $f\gg f_{\rm c}$, ${\rm PSD}_{\rm{b}} \sim f^{-2}$. Experimentally we find that for $f <\unit{100}{\hertz}$, ${\rm PSD_b} \propto f^n$ with $n = -0.02 \pm 0.05$ (low frequency regime or LF, black distribution in Fig. \ref{fgr:PSDTube}(b), N = 59) and for $f > \unit{3}{\kilo\hertz}$, $n=-1.91 \pm 0.04$ (high frequency regime or HF, N = 59).
    
    The PSD of an isolated bead only differs from the one of the same bead connected to a bare membrane nanotube in the low frequency regime, while we observe no differences for frequencies above $\unit{3}{\kilo\hertz}$ (Fig. \ref{fgr:PSDTube}(a)). Indeed, the exponent at high frequency for beads connected to a nanotube is $n_{\rm{HF}} =~$-$~1.88 \pm 0.04$ (mean $\pm$ s.e.m., N = 51), similar to the one of an isolated bead $n_{\rm{HF}} =~$-$~1.91 \pm 0.04$ (N = 59). In the low frequency regime, the power law exponents are respectively $n_{\rm{LF}} = -0.02 \pm 0.05$ for isolated beads (black distribution in Fig. \ref{fgr:PSDTube}(b)) and $n_{\rm{LF}} = -0.18 \pm 0.05$ for beads connected to nanotube (magenta distribution). 
    
    Let us now address the difference observed in the low frequency regime. We previously described the PSD of a bead connected to a membrane nanotube as \cite{Valentino2016}:
    
    \begin{equation}\label{eq:PSDV}
        {\rm PSD}_{\rm{t}}(f) = \frac{k_{\rm{B}}T}{12\pi^3\eta r_{\rm{bead}}} \frac{1+\sqrt{f_{\rm{t}}/f}}{(f_{\rm{c}}+\sqrt{f_{\rm{t}}f})^2+(f+\sqrt{f_{\rm{t}}f})^2}
    \end{equation}
    
    where $f_{\rm{c}}$ reflects the optical trap stiffness  and $f_{\rm{t}}$ is a characteristic frequency of the nanotube, given by:
    
    \begin{equation}\label{eq:ft}
        f_{\rm{t}} = \frac{2}{9}\frac{F\eta_{\rm{I}}}{(\pi\eta r_{\rm{bead}})^2}
    \end{equation}
    
    with $F$ the mean nanotube force maintenance, $\eta$ and $\eta_{\rm{I}}$ the viscosities of the surrounding and inside fluid, respectively, and $r_{\rm{bead}} = \unit{1.50}{\micro\meter}$ the radius of the bead. We assume $\eta_{\rm I}$ and $\eta$ to be the viscosity of pure water $\unit{10^{-3}}{\pascal\cdot\second}$. Here, the force is given by $F = 2\pi\sqrt{2\kappa\sigma}$, with $\kappa$ the membrane bending modulus \cite{Waugh1987}. Therefore, a decrease in membrane tension leads to a decrease in the maintenance force , which ranges in $\unit{0.2-15}{\pico\newton}$ (median $F = \unit{4}{\pico\newton}$, N = 51, Fig. \ref{fgr:ForceEvolution}). Compared to \cite{Valentino2016}, the mean nanotube force maintenance is here lower since we are at lower tension. Using our typical measured forces, equation \eqref{eq:ft} leads to an estimate of $f_{\rm t} = \unit{2-150}{\hertz}$. 
    
    The PSD of a bead connected a nanotube is described by ${\rm PSD_t}$ (equation $\eqref{eq:PSDV}$). To highlight the difference between a free bead and a bead connected to a tube, we present in Fig. \ref{fgr:PSDTube}(c) the experimental ratio $\frac{{\rm PSD_t}}{{\rm PSD_b}}(f)$ averaged on N = 51 nanotubes. These data are thus fitted by the theoretical ratio between equation $\eqref{eq:PSDV}$ and equation $\eqref{PSDfree}$ that yields:
    
    \begin{equation}\label{eq:PSDratio}
        \frac{{\rm PSD_t}}{{\rm PSD_b}}(f) = \left(1+\sqrt{\frac{f_{\rm t}}{f}}\right)\left(\left(f_{\rm c}+\sqrt{f_{\rm t}f}\right)^2+\left(f+\sqrt{f_{\rm t}f}\right)^2\right)\left(f_{\rm c}^2+f^2\right)^{-1}
    \end{equation}
    where $f_{\rm c}=\unit{320}{\hertz}$ is the mean value on N = 59 isolated beads, and $f_{\rm t}$ is a free parameter of the fitting. From this fit, we obtain $f_{\rm t} = \unit{3.1\pm3.5}{\hertz}$. This nanotube frequency is indeed between $f_{\rm t} = \unit{0}{\hertz}$ (corresponding to an isolated bead) and $f_{\rm t} \simeq \unit{220}{\hertz}$ (measured for high membrane tension nanotubes in \cite{Valentino2016}). We extract $f_{\rm t}$ from the averaged ratio obtained experimentally, that does not take into account variability of $f_{\rm c}$ (Fig. \ref{fgr:PSDTube}(a), inset). This might explain the discrepancy between our experimental value of $f_{\rm t}$ and the calculated value presented above.
    
    We conclude that the fluctuations of a bare membrane nanotube at low tension increase bead fluctuations for frequencies below $\unit{100}{\hertz}$ (Fig. \ref{fgr:PSDTube}(c)) and is captured by equation \eqref{eq:PSDV} in the range of low membrane tensions.
    
\begin{figure*}[b]
    \centering
    \includegraphics[width=\textwidth]{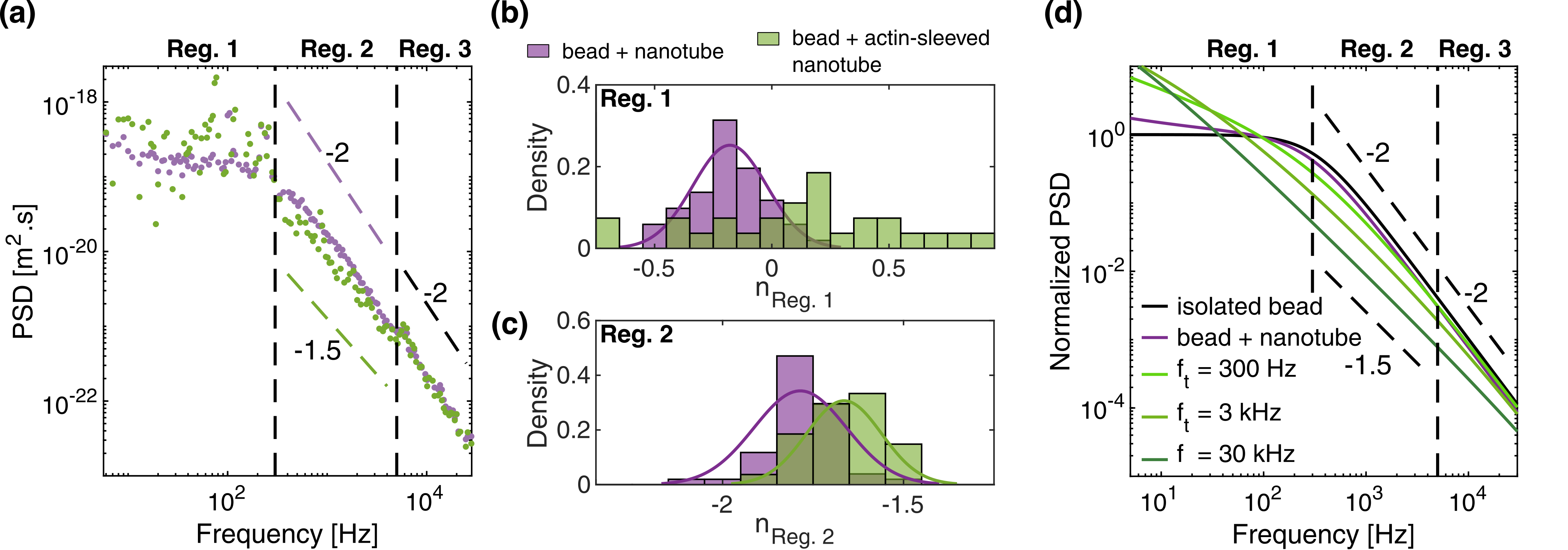}
    \caption{(a) PSDs as function of the frequency $f$ for a bare membrane nanotube (magenta) and after covering with actin (green). We divide the PSDs into three regimes: Reg. 1 for $f<\unit{300}{\hertz}$, Reg. 3 for $f>\unit{5}{\kilo\hertz}$ and Reg. 2 in between. Dashed lines indicate -2 and -1.5 slopes. (b and c) PSD exponent within \unit{[10-100]}{\hertz} range (b) and \unit{[0.5-5]}{\kilo\hertz} range (c) for bare membrane nanotubes (magenta, N = 51) and actin-sleeved membrane nanotubes (green, N = 27). p-values calculated using Students t-test. *** p $<$ 0.001. (d) Theoretical PSDs using equation \eqref{eq:PSDV} with $f_{\rm{c}} = \unit{320}{\hertz}$ and $f_{\rm{t}} = \unit{0}{\hertz}$ (black), $f_{\rm{t}} = \unit{3.1}{\hertz}$ (magenta), $f_{\rm{t}} = \unit{300}{\hertz}$ (light green), $f_{\rm{t}} = \unit{3}{\kilo\hertz}$ (intermediate green) and $f_{\rm{t}} = \unit{30}{\kilo\hertz}$ (dark green). Dashed lines indicate -2 and -1.5 slopes.}
    \label{fgr:PSDActin}
\end{figure*}

\subsection*{Fluctuations of actin-covered membrane nanotubes}

    Next we address how bead fluctuations are affected by the presence of an actin sleeve. The PSD of a bead connected to a nanotube is displayed in Fig. \ref{fgr:PSDActin}(a) in the presence (green) and the absence (magenta) of an actin sleeve. For frequencies below $f_{\rm c} \simeq \unit{300}{\hertz}$ (Reg. 1, Fig. \ref{fgr:PSDActin}(a)) and above $10\times f_{\rm c} \simeq \unit{3}{\kilo\hertz}$ (Reg. 3, Fig. \ref{fgr:PSDActin}(a)), the presence of actin does not visibly affect membrane nanotubes, whereas the intermediate regime (Reg. 2, Fig. \ref{fgr:PSDActin}(a)) exhibits differences. The region boundaries are defined as follows: Reg. 1 goes from our lowest accessible frequency, $\unit{10}{\hertz}$ to $f_{\rm c}$, Reg. 2 goes from $f_{\rm c}$ to $10\times f_{\rm c}$ to obtain a large range of frequency in the region where actin effect is apparent, and Reg. 3 goes from $10\times f_{\rm c}$ to $\unit{25}{\kilo\hertz}$, our maximal accessible frequency. We have checked that our results are not affected by the choice of these boundaries (Fig. \ref{fgr:allReg}).
    
    In Reg. 1, data are more dispersed in the presence of actin than before actin polymerization (Fig. \ref{fgr:PSDActin}(a)). The distribution of the exponent $n_{\rm Reg.~1}$ in both cases is given in Fig. \ref{fgr:PSDActin}(b). Whereas the distribution of $n_{\rm Reg.~1}$ in the absence of actin can be fitted by a gaussian, this is not the case in the presence of an actin sleeve. In Reg. 3, the exponent is similar with ($n_{\rm{Reg.~3}} = -1.87 \pm 0.04$) and without actin ($n_{\rm{Reg.~3}} = -1.88 \pm 0.04 $). 
    
    In the intermediate regime, Reg. 2, the presence of the actin sleeve affects the exponent of the PSD (Fig. \ref{fgr:PSDActin}(a and c)). We get $n_{\rm{Reg.~2}} = -1.66 \pm 0.04$ with actin (green) and a significantly lower exponent $n_{\rm{Reg.~2}} = -1.79 \pm 0.04$ without actin (magenta). We explore in Fig. \ref{fgr:allReg} the influence of regions boundaries on these exponent, and conclude that no substantial differences with the one considered here.
    
    A first attempt to explain this difference in Reg. 2 is to consider \textit{transverse} thermal fluctuations, such as the one from a guitar string, which we initially observed on membrane nanotube shapes (Fig. \ref{fgr:setup}(c)). Adapting a framework previously developed for neurite cores, surrounded by cytoskeleton and a plasma membrane \cite{Garate2015}, leads to $n_{\rm{Reg.~2}} = -1.25$ (see Appendix in Supplementary material for detailed calculations). This discrepancy shows that the \textit{transverse} fluctuations of the nanotube do not explain our data.

    Another hypothesis is that the viscoelasticity of the actin network could affect radial undulations of the nanotube. The framework recalled above (equation \eqref{eq:PSDV}) introduces a characteristic frequency $f_{\rm t}$ given by equation \eqref{eq:ft}, which is determined by the difference in viscosity between the inside and the outside of the nanotube. The bottom term $\pi\eta r_{\rm{bead}}$ catches the thermal fluctuations of the isolated bead in the surrounding viscous medium while the upper term expresses the damping of nanotube fluctuations due to the viscosity $\eta_{\rm I}$ inside the nanotube. Here, an actin sleeve of few hundreds of nanometers surrounds the membrane nanotube \cite{Allard}. We propose that this sleeve affects the membrane nanotube peristaltic modes by increasing the viscosity around the nanotube. Indeed, equation $\eqref{eq:ft}$ shows that $f_{\rm t}$ depends on the ratio between $\eta_{\rm I}$ and $\eta^2$. In the presence of an actin sleeve, we then assume that the characteristic frequency would be:
    \begin{equation}
        f_{\rm{t}} = \frac{2}{9}\frac{F\eta_{\rm{actin}}}{(\pi\eta r_{\rm{bead}})^2}
    \end{equation}
    with $\eta_{\rm{actin}}$ the viscosity outside the nanotube due to the actin sleeve. To look closely at how the PSD from equation $\eqref{eq:PSDV}$ behaves as a function of the characteristic frequency $f_{\rm t}$, we display in Fig. \ref{fgr:PSDActin}(d) the theoretical PSDs with $f_{\rm c} = \unit{320}{\hertz}$ for an isolated bead (black, $f_{\rm t} = \unit{0}{\hertz}$), a bead connected to a bare membrane nanotube (magenta, $f_{\rm t} = \unit{3.1}{\hertz}$) and for various $f_{\rm t} = \unit{0.3, 3, 30}{\kilo\hertz}$ (green). We thus capture the change from -2 to -1.5 of the PSD exponents at intermediate frequencies while increasing $f_{\rm t}$ up to $\unit{30}{\kilo\hertz}$. In the presence of an actin sleeve, we estimate $f_{\rm t} \simeq \unit{3}{\kilo\hertz}$, 3 orders of magnitude higher than the bare nanotube case (compare magenta to green curves in Figs. \ref{fgr:PSDActin}(a,d)).

    In addition, let us now consider a frequency $f$ in Reg. 2 such as $f_{\rm c} \ll f \ll f_{\rm t}$. Rewriting equation \eqref{eq:PSDV} yields:

    \begin{equation}\label{eq:Taylor}
        {\rm PSD_t}(f) \propto \frac{\sqrt{f_{\rm{t}}/f}}{f_{\rm{t}}f}\frac{1+\sqrt{f/f_{\rm{t}}}}{(1+\frac{f_{\rm{c}}}{\sqrt{f_{\rm{t}}f}})^2+(1+\frac{f}{\sqrt{f_{\rm{t}}f}})^2}
    \end{equation}
    
    where $\sqrt{f/f_{\rm{t}}}\ll1$, $\frac{f_{\rm c}}{\sqrt{f_{\rm{t}}f}}\ll1$ and $\frac{f}{\sqrt{f_{\rm{t}}f}}\ll1$. The zero order of the Taylor expansion of the second ratio in equation \eqref{eq:Taylor} gives 1 and thus leads to ${\rm PSD_t}(f) \propto \frac{1}{\sqrt{f_{\rm t}}}\frac{1}{f^{3/2}}\propto f^{-1.5}$ for \textit{peristaltic} modes, in agreement with our experimental distribution of exponents in Reg. 2 in the presence of an actin sleeve (Fig. \ref{fgr:PSDActin}(a and c)). Therefore, the viscoelasticity of the branched actin network at the surface of membrane nanotubes reduces radius undulations along the nanotube.

\section*{Conclusion}

    \textit{In vivo}, several physiological processes involve membrane nanotubes, that are highly dynamic while interacting with the actin cytoskeleton. For example, plasma membrane protrusions at the front of a cell and filled with actin bundles, called filopodia, present large spatial fluctuations over time that are dominated by bending modes \cite{zidovska2011}. Moreover, under pulling by an optical trap, the force exerted by the filopodium tip exhibits pN-range fluctuations \cite{bornschlogl2013}. In the case of endocytosis or endoplasmic reticulum remodeling, actin interacts with membrane nanotubes in the reverse geometry (compared to filopodia), and their dynamics is poorly explored. In addition, the behavior of membrane nanotube at various frequencies remains to be elucidated.
    
    Recording fluctuations is a non-invasive tool to probe the mechanics of soft objects such as membrane nanotubes. We experimentally measure the fluctuations of optically trapped beads connected to nanotubes in the range of membrane tensions measured \textit{in vivo} (\unit{0.2-50\times10^{-6}}{\newton\per\meter}). We calculate the PSD of the connected beads and show how membrane tension and actin coverage affect their fluctuations.
    
    A PSD reflects the amplitude of bead fluctuations over time, where low frequency regime corresponds to long observation time scale, and vice-versa. In this work, we introduce two time scales to describe bead fluctuations. First, for times below \unit{0.1}{\milli\second}, the fluctuations of the bead are not affected by the presence of the nanotube (Reg. 3 in Fig. \ref{fgr:PSDTube}). Second, the nanotube increases bead fluctuations at times longer than \unit{10}{\milli\second} (Fig. \ref{fgr:PSDTube}(c) and Reg. 1 in Fig. \ref{fgr:PSDActin}). These fluctuations increase with membrane tension as the bead connected to a nanotube explores a larger area inside the trap than an isolated bead. The intermediate time scale regime (Reg. 2) is sensitive to the presence of an actin sleeve that damps nanotube fluctuations (Fig. \ref{fgr:setup}(c)) and drops bead fluctuations (Fig. \ref{fgr:PSDActin}(a)). The damping of the power spectral density of an actin-coated liposome, overall frequencies, has previously been reported \cite{helfer2001}. In this case, the liposome was covered with an actin cortex while here the PSD mostly reflects the nanotube undulations.
    
    A model previously described in \cite{Valentino2016}, where squeezing modes of the nanotube influence bead fluctuations (equation \eqref{eq:PSDV}), introduces a characteristic frequency $f_{\rm t}$ (equation \eqref{eq:ft}) proportional to the force nanotube maintenance $F = 2\pi\sqrt{2\kappa\sigma}$ and the viscosity $\eta_{\rm I}$ inside the nanotube. We postulate that squeezing modes are damped by the presence of an external viscous material as an internal viscosity increase would. Therefore, the characteristic frequency, characterized by $f_{\rm t} \propto \eta_{\rm actin}\sqrt{\sigma}$, captures both the role of membrane tension and protein covering of nanotubes on bead fluctuations.
    
    An isolated bead corresponds to $f_{\rm t} = \unit{0}{\hertz}$ (in the absence of nanotube). The presence of a bare membrane nanotube connected to the bead increases $f_{\rm t}$ to \unit{3.1\pm3.5}{\hertz} at low membrane tensions (Fig. \ref{fgr:PSDTube}(c)) while we get $f_{\rm t} \simeq \unit{220}{\hertz}$ for high membrane tensions (\unit{10-200\times10^{-6}}{\newton\per\meter}, \cite{Valentino2016}).

    Equation $\eqref{eq:ft}$ provides an estimate of the viscosity of a \textit{branched} actin network at a nanometric scale: $\eta_{\rm{actin}} \simeq \unit{1}{\pascal\cdot\second}$. This measure is much larger than water viscosity, and supports our model assumptions.
    
    The viscosity of actin networks is highly dependent on the temperature and the actin concentration \cite{maruyama1974}, on the present of cross-linkers and their relative amount \cite{gardel2006,lieleg2007}, and on whether the network is in 3D \cite{Gardel2008} or coated on a membrane \cite{noding2018}, therefore the literature provides values of viscosity that are sparse ($\eta = \unit{0.2-2000}{\pascal\cdot\second}$, \cite{bausch1998,bausch1999,wottawah2005,gardel2006,lieleg2007,Gardel2008,noding2018}). Comparing our measured value to these references is hard for several reasons. First, in most of these cases, geometry is different than ours: actin is mainly coupled to flat membranes whereas our actin sheath is a hollow cylinder. Moreover, this sheath has a size close to the actin meshsize, and thus comparing its properties to bulk actin gels is difficult. Even though we do not extract a frequency-dependent value for the actin viscosity, it is worth noting that, in all references above, the rheological properties of actin networks are probed up to $\unit{100}{\hertz},$ whereas we extend the accessible frequencies up to $\unit{25}{\kilo\hertz}$.
    
    Therefore, this work unveils how the dynamics of membranous structures \textit{in vivo} are sensitive to membrane tension and cytoskeletal protein assembly in their vicinity. Inside the cell, the presence of actin could modulate nanotube radius fluctuations and thus favor the binding of nanotube remodeling proteins that ultimately lead to nanotube stability or scission \cite{Morlot2012}. In the present work, the actin viscoelasticity affect local nanotube radii that we detect thanks to the bead at the tip of the nanotube. However, the optical setup is technically designed to directly access microrheology of actin networks (or any polymer) coupled with membrane in a high frequency regime, up to $f=\unit{25}{\kilo\hertz}$, while most of classical techniques often explore the microrheology up to $\unit{100}{\hertz}$ \cite{crocker2000,furst2005,robertson2018}.

\begin{acknowledgments}
This work was supported by the French Agence Nationale pour la Recherche (ANR), grants ANR 09BLAN0283, ANR 12BSV5001401, ANR 15CE13000403, and ANR 18CE13000701, and by Fondation pour la Recherche M\'edicale,
grants DEQ20120323737 and FDT201904007966, and ERC Consolidator Grant 771201. Our groups belong to the CNRS consortium CellTiss. The authors acknowledge John Manzi for purifying the proteins.

A.A executed experiments and analyzed data. A.A., T.B. and C.C. performed the theoretical models. T.B., C.S. and C.C. designed the research. All authors contributed to writing the paper. The authors declare that they have no competing interests.
The data that support the plots within this paper and other findings of this study are available from the corresponding authors upon request.
\end{acknowledgments}

\bibliography{reference.bib}


\section*{Supplemental Material}

\subsection*{Materials and methods}
\subsubsection*{Experimental setup}

As previously described \cite{Valentino2016}, and sketched in Fig. 1(a), to record lateral fluctuations of membrane nanotube we use a custom built optical tweezer based on an infrared laser ($\lambda$ = \unit{1064}{\nano\meter}, $P$ = \unit{5}{\watt}, YLM-5-LP-SC, IPG Laser, Germany) positioned by an AOD pair (MT80-A1 \unit{51064}{\nano\meter}, AA Opto Electronic, France). The beam is imaged on the back focal plane of a water immersion objective (PLAN APO VC 60x A/1.2WI IFN 25 DIC N2, Nikon, Japan). This objective is related to a perfect focus system (PFS, Ti-ND6-PFS-MP, Nikon). The laser is coupled in the optical path of an inverted microscope (Ti-E, Nikon) by several dichroic mirrors (Beamsplitter, AHF, Germany). 

We visualize images with a spinning disk (SD) confocal microscope (CSUX1 YOKOGAWA, Andor, Ireland) and a high resolution sCMOS Camera (Andor). The setting parameters for imagery (laser power, acquisition time, optical filters) are kept constant in all cases described in this work (bare and actin-covered nanotubes). We have checked that the presence of an actin signal does not affect the lipid signal (Fig. \ref{fgr:intensity}). To extrude a membrane nanotube, we first trap a streptavidin-coated polystyrene bead (\unit{3.05}{\micro\meter} diameter, streptavidin-coated, Spherotech, Illinois, USA). We then attach to this bead a biotinylated liposome, slightly adherent to the bottom surface of the chamber. Moving the chamber with a 2D piezo stage at a constant speed (MS 2000, ASI, USA), allows us to form a nanotube between the liposome and the bead.

The trapping laser is collected in transmission by a water immersion objective (NIR APO 60x/0.8 W DIC N2, Nikon). We record the position of the bead relative to the trap center based on the back focal plane technique \cite{Gittes1998}. The interference signal between the unscattered laser light and the light scattered by the bead is imaged on a quadrant-photodiode (QPD, PDQ-30-C, Thorlabs, Germany). The signal is acquired by a data acquisition card (NI PCIe-6363, National Instruments, Austin, USA), at a rate of $\unit{250}{\kilo\hertz}$, which gives a temporal resolution of $\unit{4}{\micro\second}$. The calibration of the QPD on a bead allows us to determine the relation between the QPD voltage $V_{\rm{QPD}}$ and the distance $d$ separating the center of the bead from the center of the trap, as detailed in \cite{Valentino2016}. In our experiment, nanotube forces correspond to the bead position and is restricted to the linear region. After proper calibration, the voltage from the QPD is proportional to the bead displacement, with a typical conversion coefficient of $\unit{0.5}{\milli\volt\per\nano\meter}$. The voltage noise of the QPD is $<\unit{0.3}{\milli\volt}$, thus the spatial resolution detectable by the photodiode is about $\unit{1}{\nano\meter}$.

We synchronize instrument controlling and data recording by LabView software (National Instruments). Image acquisition is done by iQ3 software (Andor). We analyze data with Matlab software (The MathWorks, Natick, MA).

\setcounter{equation}{0}
\renewcommand{\theequation}{S\arabic{equation}}%
\setcounter{figure}{0}
\renewcommand{\thefigure}{S\arabic{figure}}%

\subsubsection*{Lipids, buffers and reagents}
We purchase lipids EPC (L-$\alpha$-phosphatidylcholine from egg yolk), DS-PE-PEG(2000)-biotin (1,2-distearoyl-sn-glycero-3-phosphoethanolamine-N [biotinyl-(polyethylene glycol) 200]) and 18:1 DGS-NTA(Ni) (1,2-dioleoyl-sn-glycero-3-[(N-(5-amino-1-carboxypentyl) iminodiacetic acid)succinyl]) from Avanti Polar Lipids (Alabaster, USA). We obtain Texas Red DHPE (1,2-dihexadecanoyl-sn-glycero-3-phosphoethanolamine, triethylammonium salt) from Thermo Fisher (Waltham, USA).
    
We purchase all chemicals from Sigma Aldrich. The internal buffer (TPI) consists of \unit{2}{\milli\mega} Tris and \unit{200}{\milli\mega} sucrose. The actin polymerization occurs in the external buffer (TPE) containing \unit{1}{\milli\mega} Tris, \unit{50}{\milli\mega} \chemform{KCl}, \unit{2}{\milli\mega} \chemform{MgCl_2}, \unit{0.1}{\milli\mega} DTT, \unit{2}{\milli\mega} ATP, \unit{0.02}{\gram\per\liter} $\beta$-casein and \unit{95}{\milli\mega} sucrose. TPE\textsubscript{inj}, limiting actin polymerization inside the micropipette, consists of \unit{1}{\milli\mega} Tris, \unit{1}{\milli\mega} \chemform{MgCl_2}, \unit{0.1}{\milli\mega} DTT, \unit{0.02}{\gram\per\liter} $\beta$-casein and \unit{195}{\milli\mega} sucrose. TPA, a high osmolarity buffer, contains \unit{1}{\milli\mega} Tris, \unit{1}{\milli\mega} \chemform{MgCl_2}, \unit{0.1}{\milli\mega} DTT, \unit{0.02}{\gram\per\liter} $\beta$-casein and \unit{395}{\milli\mega} sucrose. All buffers are adjusted at pH 7.4 and their osmolarity are set at \unit{200}{\milli osm\per\kilo\gram} (\unit{400}{\milli osm\per\kilo\gram} for TPA). We measure osmolarities with a vapor pressure osmometer (Vapro 5600, Wescor, USA). Monomeric actin is prepared in a G-buffer composed of \unit{2}{\milli\mega} Tris, \unit{0.2}{\milli\mega} \chemform{CaCl_2}, \unit{0.2}{\milli\mega} DTT, \unit{0.2}{\milli\mega} ATP (pH 8.0).

We purchase actin and the porcine Arp2/3 complex from Cytoskeleton (Denver, USA), fluorescent Alexa Fluor 488 actin conjugate (actin-488) from Molecular Probes (Eugene, USA).
Purification of mouse $\alpha 1 \beta 2$ capping protein (CP) is described elsewhere \cite{Palmgren2001}. His-pVCA-GST (pVCA, the proline rich domain-verprolin homology-central-acidic sequence from human WASP, starting at amino acid Gln150) is purified as for PRD-VCA-WAVE \cite{Havrylenko2015}. Untagged human profilin is purified as in \cite{Carvalho2013}.
A solution of \unit{30}{\micro\mega} monomeric actin containing 15\% of labelled actin-488 is obtained by incubating the actin solution in G-Buffer over two days at \unit{4}{\degree\Celsius}.
Commercial proteins are used with no further purification and all concentrations are checked by a Bradford assay.

\subsubsection*{Membrane and actin sleeve}

We will further describe membrane nanotube pulling from liposomes formed using the electroformation method \cite{Angelova1986}. The lipid mixture (molar ratio EPC/DGS-Ni/DSPE-PEG-biotin/Texas Red DHPE of 89.4/10/0.1/0.5) is aliquoted at \unit{2.5}{\gram\per\liter} in chloroform/methanol at volume ratio 5/3. A volume of \unit{5}{\micro\liter} of this solution is spread on an ITO-coated (Indium Tin Oxide) glass slide (63691610PAK, Sigma Aldrich, Germany), and dried in vacuum for \unit{2}{\hour}. We face the two conductive slides, sealed with Vitrex (Vitrex Medical A/S, Denmark), to form a chamber. We then hydrate the film with TPI and apply an oscillating electric field (\unit{10}{\hertz}, \unit{3}{\volt} peak to peak) during \unit{2}{\hour}. Liposomes are stored at \unit{4}{\degree\Celsius} for up to two weeks.

Prior to experiments, we clean and passivate the glass surfaces. We sonicate glass coverslips (\unit{0.13-0.16}{\milli\meter}, Menzel Gl\"aze, Australia) in 2-propanol for 5 minutes, extensively rinsed with water and dried under filtrated compressed air. Then the glass surfaces are activated by a plasma cleaner (PDC-32G, Harrick Plasma, USA) during 2 minutes, followed by a 30 minutes passivation using \unit{0.1}{\gram\per\liter} PLL(20)-g[3.5]-PEG(2) (SuSos, Switzerland) in a \unit{10}{\milli\mega} Hepes solution (pH 7.4). We assemble the experimental chamber facing two glass coverslips separated by a \unit{1}{\milli\meter} steel spacer. The chamber is filled with a \unit{100}{\micro\liter} solution, diluted in TPE and containing \unit{3}{\micro\mega} profilin, \unit{37}{\nano\mega} Arp2/3 complex, \unit{25}{\nano\mega} CP, \unit{2}{\micro\liter} liposomes in TPI, and \unit{1}{\micro\liter} polystyrene beads diluted 100 times in TPE. 
    
Micropipettes are prepared from borosilicate capillaries (\unit{0.7{\milli\meter}/ 1.0}{\milli\meter} for inner/outer diameter, Harvard Apparatus, USA), using a puller (P2000, Sutter Instrument, USA) with parameters previously described in \cite{Valentino2016}. Micropipette tips are then micro-forged (MF 830, Narishige, Japan) up to an internal diameter of \unit{10}{\micro\meter}. Micropipettes are filled by aspirating \unit{1}{\micro\liter} of the desired solution. Mineral oil is filled on the other side of the micropipette using a MicroFil (\unit{250}{\micro\meter} ID \unit{350}{\micro\meter} OD \unit{97}{\milli\meter} long, World Precision Instrument, UK). We prepare two micropipettes: the first one contains \unit{2}{\micro\mega} pVCA, \unit{0.01}{\gram\per\liter} sulforhodamine-B (to monitor the microinjection), in TPE; the second one contains \unit{3}{\micro\mega} actin-488 and \unit{3}{\micro\mega} profilin, in TPE\textsubscript{inj}, adjusted to the osmolarity of \unit{200}{\milli Osm \per\kilo\gram} with TPA.

Note that profilin is present in the actin microinjection pipette and in the chamber, so that actin polymerization is prevented in the micropipette and in solution, and occurs mainly at the membrane surface.

Each micropipette is set up into the chamber, and connected to two separated reservoirs to control independently the injection pressures. The chamber is sealed on each side by adding mineral oil, to block evaporation over the time of the experiment.

\subsection*{Appendix}
\subsubsection*{Power spectral density calculations}

We first record the position $d$ of the trapped bead relative to the center of the trap as a function of time. Using fast Fourier transformation ($FFT$) we infer the power spectral density as a function of the frequency $f$:
    
\begin{equation}
    {\rm PSD}(f) = \frac{FFT(d)\times FFT^*(d)}{T_{\rm{exp}}}
    \label{PSDdef}
\end{equation}
    
where $FFT^*$ is the conjugate of $FFT$ and $T_{\rm exp}$ the time of the experiment. Power spectral densities PSD as function of the frequency $f$ are generated using the FFT algorithm. Power laws calculation is performed on logarithmic transformation of experimental PSD. The exponent $n$ for each regime is then deduced from a linear fit: $\log_{10}({\rm PSD}) = n\times\log_{10}(f) + a$. This method reduces the computational error in the exponent $n$ calculation. We display $n$ as mean $\pm$ s.e.m..

\subsubsection*{Transverse mode fluctuations}

The PSD of a fluctuating bead connected to a nanotube reflects thermal fluctuations of the bead itself in parallel with membrane nanotube fluctuation transmitted to the bead. We describe in the main text the fluctuations induced by \textit{peristaltic} undulations. We here focus on \textit{transverse} modes of a nanotube of length $L$ as described in \cite{Garate2015} for neurite cores, surrounded by cytoskeleton and a plasma membrane, a composite system characterized by an axial tension $\mathcal{T}$ and a bending flexural rigidity $\mathcal{B}$. Decomposing into Fourier modes with amplitudes $h_j$ and wave vectors $q_j$ yields: $|h_j|^2 = \frac{k_{\rm{B}}T}{L(\mathcal{B}q_j^4+\mathcal{T} q_j^2)}$ \cite{Garate2015}. Moreover the dispersion relation is given by $\omega(q) = \frac{\mathcal{T} q^2 + \mathcal{B} q^4}{\eta_{\rm{e}}}$  where $\eta_{\rm e}$ is the effective dissipation \cite{Betz2009,Garate2015}.

In the present case, \textit{transverse} modes of the nanotube shift the bead of a relative displacement $\delta l(t) = \int_0^L |\partial_xh(x,t)|dx = \sum_j \delta l_j e^{-i\omega_jt}$. Adapting the calculation from \cite{Garate2015} with this $\delta l$ constrain gives the theoretical expression for the PSD:

\begin{equation}\label{eq:PSDG}
{\rm PSD}(\omega)  =  \frac{\eta_{\rm{e}}k_{\rm{B}}T}{\pi}\int \frac{L^2q^2}{(\mathcal{B}q^4+\mathcal{T}q^2)^2+(\eta_{\rm{e}}\omega)^2}dq
\end{equation}

In the case where the cytoskeletal bending is dominant ($\mathcal{B}q^4\gg\mathcal{T}q^2$), equation \eqref{eq:PSDG} reads ${\rm PSD} \propto \int \frac{dq}{q^6}\propto q^{-5}$ and the dispersion relation becomes $2\pi f=\omega \propto q^4$. Altogether \textit{transverse} modes yields ${\rm PSD} \propto f^{-5/4}=f^{-1.25}$.

\subsection*{Movie}

\noindent  FIG. Movie S1. Confocal images of the three bare nanotubes exemplified in Fig. 1(c): (a) crosses (b) circles and (c) stars. Acquired at a rate of one frame per second. Scale bar: \unit{20}{\micro\meter}.

\subsection*{Figures}

\clearpage

\begin{figure}
    \centering
        \includegraphics[width=100mm]{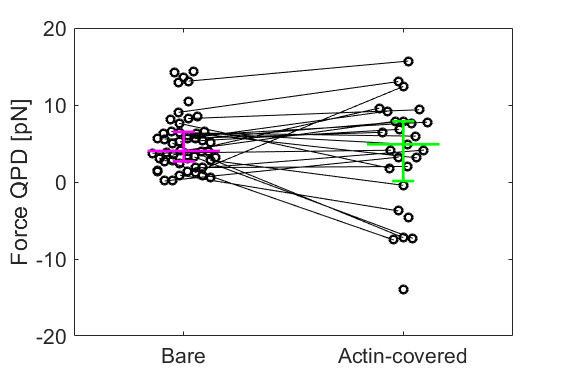}
        \caption{\textbf{Nanotube force distribution.} Distribution of the nanotube force $F_0$ ($L_{\rm tube} \simeq \unit{10-20}{\micro\meter}$) before (magenta) and after (green) actin polymerization. Bars represent respectively first, second and third quartiles. When recorded on the same nanotube, lines display the evolution of the force.}
    \label{fgr:ForceEvolution}
\end{figure}

\clearpage

\begin{figure}
    \centering
        \includegraphics[width=\textwidth]{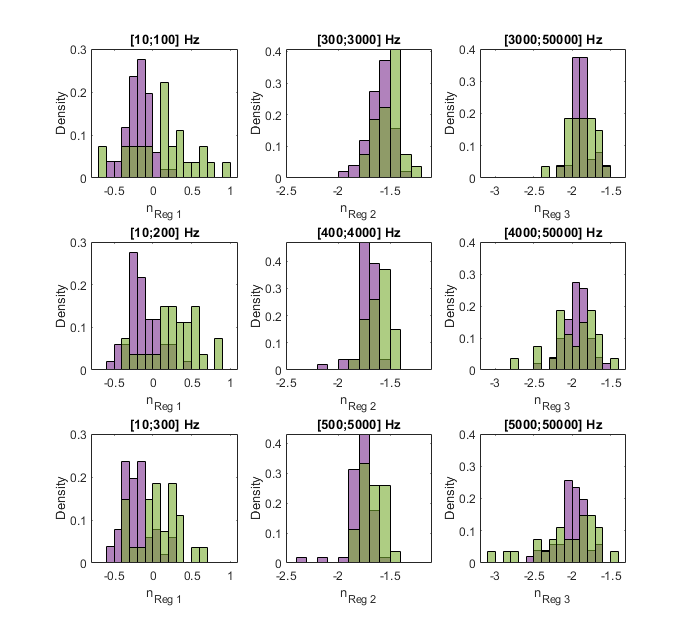}
        \caption{\textbf{Influence of the boundaries on spectral exponents.} Distribution of the spectral exponents as a function of the fitting range, from all PSD of N = 51 membrane nanotubes (magenta) and N = 27 membrane nanotubes sleeved by an actin network (green).}
    \label{fgr:allReg}
\end{figure}

\clearpage

\begin{figure}
    \centering
        \includegraphics[width=\textwidth]{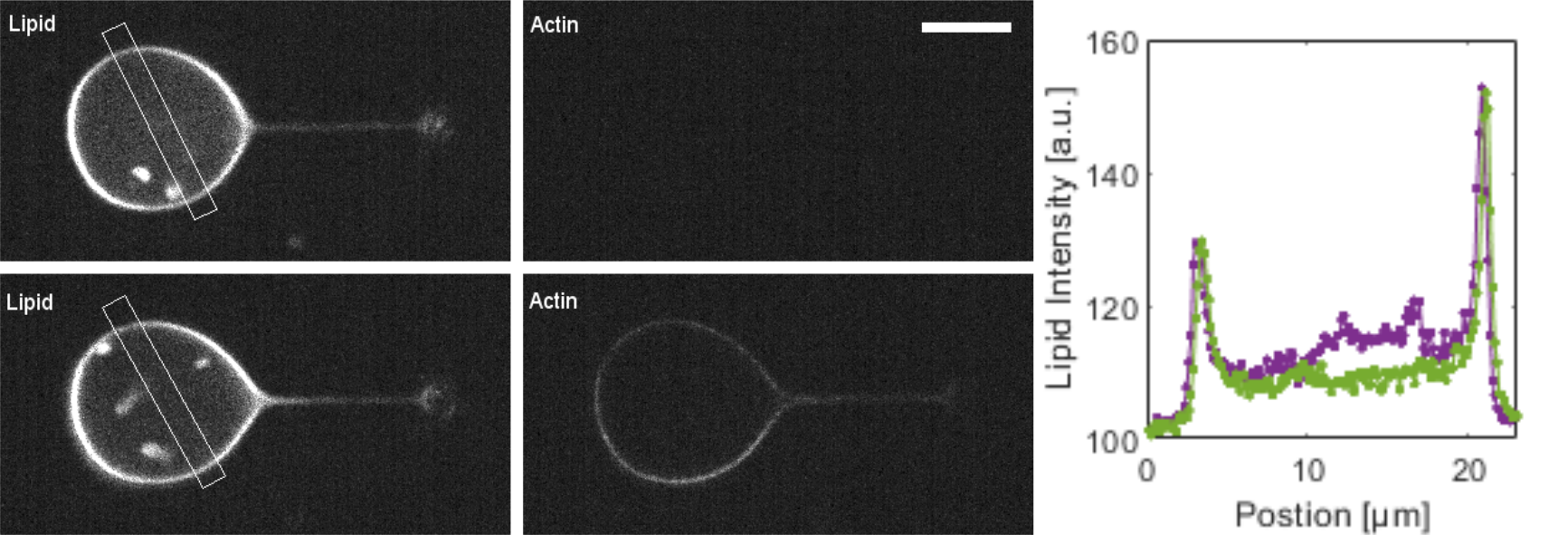}
        \caption{Confocal images of a liposome before (top) and after (bottom) actin polymerization. The plot represents the lipid channel intensity before (magenta) and after (green) actin polymerization. The intensities of the two peaks remain unchanged. Scale bar: \unit{10}{\micro\meter}.}
    \label{fgr:intensity}
\end{figure}

\end{document}